# Electro-optic properties from *ab initio* calculations in two-dimensional materials

Zhijun Jiang,[1,2,*] Hongjun Xiang,[3] Laurent Bellaiche,[4] and Charles Paillard[4,5,†]

[1] MOE Key Laboratory for Nonequilibrium Synthesis and Modulation of Condensed Matter, School of Physics, Xi'an Jiaotong University, Xi'an 710049, China
[2] State Key Laboratory of Surface Physics and Department of Physics, Fudan University, Shanghai 200433, China
[3] Key Laboratory of Computational Physical Sciences (Ministry of Education), Institute of Computational Physical Sciences, State Key Laboratory of Surface Physics, and Department of Physics, Fudan University, Shanghai 200433, China
[4] Physics Department and Institute for Nanoscience and Engineering, University of Arkansas, Fayetteville, Arkansas 72701, USA
[5] Université Paris-Saclay, CentraleSupélec, CNRS, Laboratoire SPMS, 91190, Gif-sur-Yvette, France
*zjjiang@xjtu.edu.cn
†charles.paillard@centralesupelec.fr

Electro-optic (EO) effects relate the change of optical constants by low-frequency electric fields. Thanks to the advent of Density Functional Perturbation Theory (DFPT), the EO properties of bulk three-dimensional (3D) materials can now be calculated in an *ab initio* way. However, the use of periodic boundary conditions in most Density Functional Theory codes imposes to simulate two-dimensional (2D) materials using slabs surrounded by a large layer of vacuum. The EO coefficients predicted from such calculations, if not rescaled properly, can severely deviate from the real EO properties of 2D materials. The present work discusses the issue and introduces the rescaling relationships allowing to recover the true EO properties.

## I. INTRODUCTION

Electro-optic materials are key components of modern optical technologies. Electro-optic modulators allow to modulate optical signal with a voltage input and are involved in present and future optical technologies such as optical resonators[1,2], optoelectronic oscillators[3], Mach-Zender modulators[4] or one-photon sources for quantum cryptography[5,6]. Linear EO effects, sometimes referred to as Pockels effects, relate the change of optical index to the first-order change in low-frequency electric field (up to 100 MHz). It is typically characterized by a linear EO coefficient $r_{ijk}$ such that[7]

$$\Delta\beta_{ij} = \sum_{k=1}^{3} r_{ijk}\Delta E_k \qquad (1)$$

where $\beta = \varepsilon^{-1}$ is the dielectric stiffness tensor (inverse of the dielectric permittivity tensor $\varepsilon$), and $\Delta E_k$ is the change in low-frequency applied electric field in the cartesian direction $k$.

A little more than a decade ago, the development of Density Functional Perturbation Theory (DFPT), an extension of Density Functional Theory (DFT) to linear and non-linear response properties, deepened the field of EO materials[8–10]. Thanks to the better understanding of transverse optical phonon contributions to the EO response, new strategies using strain engineered ferroelectric thin films[11] or superlattices[12] were recently devised to improve EO responses. Such avenue has also been pursued experimentally in ferroelectrics[13]. Since dimensionality reduction is a heavy trend in the electronics and photonics industry, it seems only logical that the next step in the search for efficient EO materials is to investigate 2D materials. Already, 2D materials such as graphene have proved promising to achieve EO modulators with large bandwidth[14].

2D functional materials have also attracted a lot of attention from the DFT modelling community. Beyond graphene, a large variety of 2D materials have been shown to exhibit functional properties such as thermoelectricity[15], ferroelectricity and piezoelectricity[16–19], photostriction[20], in part thanks to the contribution of *ab-initio* methods. Most popular DFT software devoted to condensed matter systems, such as Abinit[21], VASP[22–24] or Quantum Espresso[25,26], employ 3D periodic boundary conditions. These are not well tailored to the study of 2D materials. In fact, many DFT studies simulate a 2D material as a slab in a supercell (denoted as SC) containing a large amount of vacuum (see Fig. 1).

The latter allows one to get good structural and electronic properties in standard DFT calculations, if enough vacuum is added to limit the interaction between the periodic images of the 2D layer introduced by periodic boundary conditions. It is, however, not so straightforward to then determine the response functions of the 2D materials. It is thus legitimate and timely to wonder if EO coefficients obtained from DFT calculations on slabs need to be corrected in order to correspond to those of true 2D layers.

The aim of this article is to prove that such correction is indeed required. We first indicate the renormalization to be done for several EO coefficients and for several point groups. We then numerically confirm such renormalization by conducting DFT calculations. We finally summarize and discuss the impacts of the present work.

## II. METHODS

We employ here the ABINIT package[27] with the generalized gradient approximation in the form of Perdew-Burke-Ernzerhof (PBE) exchange-correlation functional[28] with norm-conserving pseudopotentials[29] to get an accurate ground state structure but used the local density approximation (LDA) to compute the electro-optic coefficients because such computation is only implemented within LDA in ABINIT[9,21,27,30]. We chose a 10×10×1 grid of special *k*-points and a plane-wave cutoff energy of 50 hartrees. The geometries were fully optimized until the force acting on each atom is smaller than 1×10⁻⁶ hartree/bohr.

## III. RESULTS AND DISCUSSION

### A. Derivation of EO coefficients of 2D layers

Let us first concentrate on the dielectric constant and span the plane inside which the 2D material lies by the Cartesian axes 1 and 2. The vacuum and 2D materials are thus in series along the third Cartesian direction (see Fig. 1). Let us now apply an electric field along the out-of-plane direction, $E_{SC,3}$. Because the 2D material is electrically in series with the vacuum in the out-of-plane direction, it effectively feels an electric field
$$tE_{2D,3} = cE_{SC,3} - (c-t)E_{vacuum,3} \qquad (2)$$
where $E_{vacuum,3}$ is the electric field in the vacuum layer along the out-of-plane direction, $c$ and $t$ are the supercell lattice constant in the third direction and the effective thickness of the 2D layer respectively. If the 2D material is insulating, one can safely assume that charges are not free. The out-of-plane electrical displacement is then continuous, and the change in electrical displacement induced by the change in electric field is the same for the whole supercell, the vacuum layer, and the 2D material:

$$D_{2D,3} = D_{vacuum,3} = D_{SC,3}. \qquad (3)$$

The resulting dielectric constant of the 2D material in the out-of-plane direction $\varepsilon_{2D,3} = \frac{\Delta D_{2D,3}}{\varepsilon_0 \Delta E_{2D,3}}$ thus differs from that of the supercell $\varepsilon_{SC,3} = \frac{\Delta D_{SC,3}}{\varepsilon_0 \Delta E_{SC,3}}$. It is important to realize that the latter, rather than $\varepsilon_{2D,3}$, is the quantity returned by standard DFPT applied to the supercell depicted in Fig. 1.

Similarly, along the in-plane directions, the 2D and vacuum layers are electrically in parallel. If follows that along the first and second directions,

$$E_{2D,i} = E_{SC,i} = E_{vacuum,i} \qquad (4)$$
$$tD_{2D,i} = cD_{SC,i} - (c-t)D_{vacuum,i} \qquad (5)$$

where i= 1 or 2.

Since DFPT only returns the dielectric constant of the supercell (i.e. 2D layer plus vacuum), one needs to rescale it so as to return the real dielectric constant of the 2D layer. It turns out that the rescaling of the dielectric constant, which relates $\varepsilon_{2D,3}$ to $\varepsilon_{SC,3}$ has already been worked out by Laturia *et al.*[31]. We recall the results below:

$$\varepsilon_{2D,i} = 1 + \frac{c}{t}(\varepsilon_{SC,i} - 1) \qquad (6)$$

$$\varepsilon_{2D,3} = \left[1 + \frac{c}{t}(\varepsilon_{SC,3}^{-1} - 1)\right]^{-1} \qquad (7)$$

where $i = 1$ or 2.

There is some degree of arbitrariness as to what the effective thickness of the 2D material is. An efficient solution is to consider the thickness based on the van der Waals bond length indicated in Ref.[31]. Specifically, we consider the van der Waals interaction and relax the n+1-layer structure, with the distance between the center of the top and bottom layer being used as the thickness of the n-layer.

Since the dielectric constant, which is a second derivative of the energy, requires renormalization in the case of 2D materials simulated with periodic boundary conditions, it is only natural to expect that third derivatives of the energy such as the electro-optic constants will be renormalized as well. In the following, we derive the renormalization relations linking the electro-optic constant calculated using DFPT in a slab geometry (denoted with an "SC" superscript, for SuperCell) and the real electro-optic constants of the 2D material.

The EO tensor of the system comprising the vacuum and the 2D layer system is non vanishing only when the point group symmetry of the supercell pertains to the following list[32]: 1 (triclinic), 2, $m$ (monoclinic), $mm2$, 222 (orthorhombic), 4, $\bar{4}$, 422, $4mm$, $\bar{4}2m$ (tetragonal), 23, $\bar{4}3m$ (cubic), 3, 32, $3m$ (trigonal), 6, $6mm$, 622, $\bar{6}$, $\bar{6}m2$ (hexagonal). Below, we show the method and derive the rescaling laws for the orthorhombic, tetragonal, hexagonal, trigonal and cubic point groups, with representative materials examples. Similar methods can be used to derive the more complicated cases of triclinic and monoclinic point groups.

## B. Orthorhombic point groups

### Point group $mm2$

We choose a setting in which the 2-fold polar axis is located along the first Cartesian direction. This is representative of many calculations of 2D materials with in-plane electrical polarization, such as SnS[33]. In this setting, the electro-optic tensor can be written using Voigt notation as[33]:

$$r_{ijk} = \begin{pmatrix} r_{11} & r_{12} & r_{13} & 0 & 0 & 0 \\ 0 & 0 & 0 & 0 & 0 & r_{26} \\ 0 & 0 & 0 & 0 & r_{35} & 0 \end{pmatrix}. \tag{8}$$

Similarly, the dielectric stiffness tensor and dielectric permittivity tensor are diagonal:

$$\varepsilon = \begin{pmatrix} \varepsilon_{11} & 0 & 0 \\ 0 & \varepsilon_{22} & 0 \\ 0 & 0 & \varepsilon_{33} \end{pmatrix} \tag{9}$$

$$\beta = \begin{pmatrix} \beta_{11} & 0 & 0 \\ 0 & \beta_{22} & 0 \\ 0 & 0 & \beta_{33} \end{pmatrix} \tag{10}$$

Let us now derive the rescaling law for the EO coefficients, *i.e.*, derive the transformation to obtain the true EO constants of the 2D layer, $r_{ijk}^{2D}$, as a function of these calculated by DFPT in a supercell, $r_{ijk}^{SC}$. We start by focusing on coefficient $r_{111}^{2D} = r_{11}^{2D}$ in Voigt's notation[7]. The linear (Pockels) electro-optic coefficients are related to the variation of the dielectric stiffness tensor under an applied electric field:

$$r_{ijk} = \frac{\Delta\beta_{jk}}{\Delta E_i} \tag{11}$$

In the case of $r_{11} = \frac{\Delta\beta_{11}}{\Delta E_1}$, we can take full advantage of the relation $\beta.\varepsilon = 1$, where 1 is the 3 × 3 identity matrix. After differentiation, it becomes

$$\Delta\beta_{ij} = -\beta_{ik}\Delta\varepsilon_{kl}\beta_{lj} \tag{12}$$

Note that Eqs. (11) and (12) are, in fact, general equations which do not depend on the point group considered. In the special case of the $mm2$ point group, we find that

$$\Delta\beta_{11} = -\frac{\Delta\varepsilon_{11}}{\varepsilon_{11}^2} \tag{13}$$

As a result, we can now write:

$$r_{11}^{2D} = \frac{\Delta\beta_{11}^{2D}}{\Delta E_1^{2D}} = -\frac{1}{\varepsilon_{11}^{2D\,2}}\frac{\Delta\varepsilon_{11}^{2D}}{\Delta E_1^{2D}} \tag{14}$$

Using Laturia's derivations[31], which we recalled in Eq. (6), and the fact that $\Delta E_1^{2D} = \Delta E_1^{SC}$ (see Eq. (4)), we find that

$$r_{11}^{2D} = \frac{\Delta\beta_{11}^{2D}}{\Delta E_1^{2D}} = -\frac{1}{\varepsilon_{11}^{2D\,2}}\frac{c}{t}\frac{\Delta\varepsilon_{11}^{SC}}{\Delta E_1^{SC}} \tag{15}$$

Finally, noting that $r_{11}^{SC} = -\frac{\Delta\beta_{11}^{SC}}{\Delta E_1^{SC}} = -\frac{1}{\varepsilon_{11}^{SC\,2}}\frac{\Delta\varepsilon_{11}^{SC}}{\Delta E_1^{SC}}$, we obtain the final rescaling law:

$$r_{11}^{2D} = \frac{c}{t}\left(\frac{\varepsilon_{11}^{SC}}{\varepsilon_{11}^{2D}}\right)^2 r_{11}^{SC} \tag{16}$$

We have expressed the renormalization procedure in terms of the SC and 2D layer dielectric constants because of its elegance. One must bear in mind that the two quantities are linked through Eq. (6), and that such latter equation then allows to obtain the 2D dielectric constant from the SC one – with this latter being the one returned by standard DFPT applied to the supercell.

Using a similar procedure, we find that

$$r_{12}^{2D} = \frac{c}{t}\left(\frac{\varepsilon_{22}^{SC}}{\varepsilon_{22}^{2D}}\right)^2 r_{12}^{SC} \tag{17}$$

The case of the $r_{31}^{2D}$ coefficient is perhaps the simplest. Equation (7) can be casted in terms of the dielectric impermeability tensor,

$$\beta_{33}^{2D} = 1 + \frac{c}{t}(\beta_{33}^{SC} - 1) \tag{18}$$

From the latter, and Eq. (4), we easily derive that

$$r_{13}^{2D} = \frac{c}{t} r_{13}^{SC} \tag{19}$$

Let us now derive the $r_{35}$ renormalization relation. From Eqs. (10) and (12), we have that

$$\Delta\beta_{31} = -\frac{\Delta\varepsilon_{31}}{\varepsilon_{11}\varepsilon_{33}} \tag{20}$$

In particular, $\Delta\varepsilon_{31}^{2D} = \frac{\Delta D_3^{2D}}{\Delta E_1^{2D}} = \frac{\Delta D_3^{SC}}{\Delta E_1^{SC}} = \Delta\varepsilon_{31}^{SC}$, which results in $\Delta\beta_{31}^{2D} = \frac{\varepsilon_{11}^{SC}\varepsilon_{33}^{SC}}{\varepsilon_{11}^{2D}\varepsilon_{33}^{2D}}\Delta\beta_{31}^{SC}$. Hence, one can write that

$$r_{35}^{2D} = \frac{\varepsilon_{11}^{SC}\varepsilon_{33}^{SC}}{\varepsilon_{11}^{2D}\varepsilon_{33}^{2D}} \frac{\Delta\beta_{31}^{SC}}{\Delta E_3^{2D}} = \frac{\varepsilon_{11}^{SC}\varepsilon_{33}^{SC}}{\varepsilon_{11}^{2D}\varepsilon_{33}^{2D}} \frac{\Delta E_3^{SC}}{\Delta E_3^{2D}} r_{35}^{SC} \tag{21}$$

Now, $\frac{\Delta E_3^{SC}}{\Delta E_3^{2D}} = \frac{\Delta E_3^{SC}}{\Delta D_3^{2D}} \frac{\Delta D_3^{2D}}{\Delta E_3^{2D}} = \frac{\Delta E_3^{SC}}{\Delta D_3^{SC}} \frac{\Delta D_3^{2D}}{\Delta E_3^{2D}} = \frac{\varepsilon_{33}^{2D}}{\varepsilon_{33}^{SC}}$ since $D_3^{2D} = D_3^{SC}$ according to Eq. (3). Eventually, we obtain the rescaling relation[1]

$$r_{35}^{2D} = \frac{\varepsilon_{11}^{SC}}{\varepsilon_{11}^{2D}} r_{35}^{SC} \tag{22}$$

Following a similar procedure, we find that the last rescaling relation is

$$r_{26}^{2D} = \frac{c}{t} \frac{\varepsilon_{11}^{SC}\varepsilon_{22}^{SC}}{\varepsilon_{11}^{2D}\varepsilon_{22}^{2D}} r_{26}^{SC} \tag{23}$$

*Point group* 222

The EO tensor in the 222 point group is of the form[32]

$$r_{ijk} = \begin{pmatrix} 0 & 0 & 0 & r_{14} & 0 & 0 \\ 0 & 0 & 0 & 0 & r_{25} & 0 \\ 0 & 0 & 0 & 0 & 0 & r_{36} \end{pmatrix}. \tag{24}$$

with the dielectric permittivity and impermeability tensors being

$$\varepsilon = \begin{pmatrix} \varepsilon_{11} & 0 & 0 \\ 0 & \varepsilon_{22} & 0 \\ 0 & 0 & \varepsilon_{33} \end{pmatrix} \tag{25}$$

$$\beta = \begin{pmatrix} \beta_{11} & 0 & 0 \\ 0 & \beta_{22} & 0 \\ 0 & 0 & \beta_{33} \end{pmatrix} \tag{26}$$

From the previous derivations, we easily obtain that

$$r_{14}^{2D} = \frac{\varepsilon_{22}^{SC}\varepsilon_{33}^{SC}}{\varepsilon_{22}^{2D}\varepsilon_{33}^{2D}} r_{14}^{SC} \tag{27}$$

$$r_{25}^{2D} = \frac{\varepsilon_{11}^{SC}\varepsilon_{33}^{SC}}{\varepsilon_{11}^{2D}\varepsilon_{33}^{2D}} r_{25}^{SC} \tag{28}$$

---

[1] The present relation is at variance with the one derived in the Supplementary Material of Ref.[33], by a factor $\frac{\varepsilon_{33}^{SC}}{\varepsilon_{33}^{2D}}$ because the derivative $\frac{\Delta E_3^{SC}}{\Delta E_3^{2D}}$ was incorrect. The coefficient $r_{35}^{2D}$ is nonetheless very small in both cases for some systems (see Ref.[33] and DFT results in the present article). The present formula should be used.

$$r_{36}^{2D} = \frac{c}{t} \frac{\varepsilon_{33}^{2D}}{\varepsilon_{33}^{SC}} \frac{\varepsilon_{11}^{SC} \varepsilon_{22}^{SC}}{\varepsilon_{11}^{2D} \varepsilon_{22}^{2D}} r_{36}^{SC} \tag{29}$$

## C. Tetragonal point groups

*Point group* $4mm$

We consider that the high-symmetry 4-fold axis lies along the third, out-of-plane, Cartesian direction. In that case, the EO tensor writes

$$r_{ijk} = \begin{pmatrix} 0 & 0 & 0 & 0 & r_{24} & 0 \\ 0 & 0 & 0 & r_{24} & 0 & 0 \\ r_{31} & r_{31} & r_{33} & 0 & 0 & 0 \end{pmatrix}, \tag{30}$$

while the dielectric tensors $\beta$ and $\varepsilon$ are

$$\varepsilon = \begin{pmatrix} \varepsilon_{11} & 0 & 0 \\ 0 & \varepsilon_{11} & 0 \\ 0 & 0 & \varepsilon_{33} \end{pmatrix} \tag{31}$$

$$\beta = \begin{pmatrix} \beta_{11} & 0 & 0 \\ 0 & \beta_{11} & 0 \\ 0 & 0 & \beta_{33} \end{pmatrix} \tag{32}$$

We obtain the following rescaling laws:

$$r_{31}^{2D} = \frac{c}{t} \frac{\varepsilon_{33}^{2D}}{\varepsilon_{33}^{SC}} \left( \frac{\varepsilon_{11}^{SC}}{\varepsilon_{11}^{2D}} \right)^2 r_{31}^{SC} \tag{33}$$

From Eq. (18), we obtain

$$r_{33}^{2D} = \frac{c}{t} \frac{\varepsilon_{33}^{2D}}{\varepsilon_{33}^{SC}} r_{33}^{SC} \tag{34}$$

At last,

$$r_{24}^{2D} = \frac{\varepsilon_{22}^{SC} \varepsilon_{33}^{SC}}{\varepsilon_{22}^{2D} \varepsilon_{33}^{2D}} r_{24}^{SC} \tag{35}$$

*Point group* $\bar{4}2m$

We assume that the improper rotation $\bar{4}$ is along the third Cartesian direction. The EO tensor has the symmetry[32]

$$r_{ijk} = \begin{pmatrix} 0 & 0 & 0 & r_{14} & 0 & 0 \\ 0 & 0 & 0 & 0 & r_{14} & 0 \\ 0 & 0 & 0 & 0 & 0 & r_{36} \end{pmatrix}, \tag{36}$$

with dielectric tensors like those in Eqs. (31) and (32). The rescaling relations are now

$$r_{14}^{2D} = \frac{\varepsilon_{11}^{SC} \varepsilon_{33}^{SC}}{\varepsilon_{11}^{2D} \varepsilon_{33}^{2D}} r_{14}^{SC} \tag{37}$$

$$r_{36}^{2D} = \frac{c}{t} \frac{\varepsilon_{33}^{2D}}{\varepsilon_{33}^{SC}} \frac{\varepsilon_{11}^{SC} \varepsilon_{22}^{SC}}{\varepsilon_{11}^{2D} \varepsilon_{22}^{2D}} r_{36}^{SC} \tag{38}$$

*Point group* $422$

The 4-fold axis is chosen to be directed along the third Cartesian axis of the supercell. The EO tensor has the symmetry[32]

$$r_{ijk} = \begin{pmatrix} 0 & 0 & 0 & r_{14} & 0 & 0 \\ 0 & 0 & 0 & 0 & -r_{14} & 0 \\ 0 & 0 & 0 & 0 & 0 & 0 \end{pmatrix}, \tag{39}$$

The rescaling law for the EO coefficient is

$$r_{14}^{2D} = \frac{\varepsilon_{11}^{SC} \varepsilon_{33}^{SC}}{\varepsilon_{11}^{2D} \varepsilon_{33}^{2D}} r_{14}^{SC} \tag{40}$$

*Point group $\bar{4}$*

The 4-fold axis is also taken to be directed along the third Cartesian axis of the supercell. The EO tensor has the symmetry[32]

$$r_{ijk} = \begin{pmatrix} 0 & 0 & 0 & r_{14} & -r_{24} & 0 \\ 0 & 0 & 0 & r_{24} & r_{14} & 0 \\ r_{31} & -r_{31} & 0 & 0 & 0 & r_{36} \end{pmatrix}, \quad (41)$$

The renormalization of the 2D EO coefficients from the SC calculated EO constants is

$$r_{31}^{2D} = \frac{c}{t} \frac{\varepsilon_{33}^{2D}}{\varepsilon_{33}^{SC}} \left(\frac{\varepsilon_{11}^{SC}}{\varepsilon_{11}^{2D}}\right)^2 r_{31}^{SC} \quad (42)$$

$$r_{14}^{2D} = \frac{\varepsilon_{11}^{SC} \varepsilon_{33}^{SC}}{\varepsilon_{11}^{2D} \varepsilon_{33}^{2D}} r_{14}^{SC} \quad (43)$$

$$r_{24}^{2D} = \frac{\varepsilon_{22}^{SC} \varepsilon_{33}^{SC}}{\varepsilon_{22}^{2D} \varepsilon_{33}^{2D}} r_{24}^{SC} \quad (44)$$

$$r_{36}^{2D} = \frac{c}{t} \frac{\varepsilon_{33}^{2D} \varepsilon_{11}^{SC} \varepsilon_{22}^{SC}}{\varepsilon_{33}^{SC} \varepsilon_{11}^{2D} \varepsilon_{22}^{2D}} r_{36}^{SC} \quad (45)$$

*Point group 4*

The 4-fold axis continues to be directed along the third Cartesian axis of the supercell. The EO tensor has the symmetry[32]

$$r_{ijk} = \begin{pmatrix} 0 & 0 & 0 & r_{14} & r_{24} & 0 \\ 0 & 0 & 0 & r_{24} & -r_{14} & 0 \\ r_{31} & r_{31} & r_{33} & 0 & 0 & 0 \end{pmatrix}, \quad (46)$$

The renormalization of the EO constants is operated by the following equations:

$$r_{31}^{2D} = \frac{c}{t} \frac{\varepsilon_{33}^{2D}}{\varepsilon_{33}^{SC}} \left(\frac{\varepsilon_{11}^{SC}}{\varepsilon_{11}^{2D}}\right)^2 r_{31}^{SC} \quad (47)$$

$$r_{14}^{2D} = \frac{\varepsilon_{11}^{SC} \varepsilon_{33}^{SC}}{\varepsilon_{11}^{2D} \varepsilon_{33}^{2D}} r_{14}^{SC} \quad (48)$$

$$r_{24}^{2D} = \frac{\varepsilon_{22}^{SC} \varepsilon_{33}^{SC}}{\varepsilon_{22}^{2D} \varepsilon_{33}^{2D}} r_{24}^{SC} \quad (49)$$

$$r_{33}^{2D} = \frac{c}{t} \frac{\varepsilon_{33}^{2D}}{\varepsilon_{33}^{SC}} r_{33}^{SC} \quad (50)$$

## D. Hexagonal point groups

*Point group $\bar{6}m2$*

In that case, in the standard orientation[7] (for which the sixfold axis is along the third Cartesian direction), the electro-optic tensor symmetry is

$$r_{ijk} = \begin{pmatrix} 0 & 0 & 0 & 0 & 0 & -r_{22} \\ -r_{22} & r_{22} & 0 & 0 & 0 & 0 \\ 0 & 0 & 0 & 0 & 0 & 0 \end{pmatrix}, \quad (51)$$

One can obtain the "real" electro-optic coefficient of the 2D layer from those calculated from standard DFPT, as implemented in a 3D periodic software such as Abinit, by applying the following rescaling:

$$r_{22}^{2D} = \frac{\Delta \beta_{22}^{2D}}{\Delta E_2^{2D}} = \frac{c}{t} \left(\frac{\varepsilon_{22}^{SC}}{\varepsilon_{22}^{2D}}\right)^2 r_{22}^{SC} \quad (52)$$

*Point group $\bar{6}$*

We again consider the case in which the high-symmetry axis points along the third Cartesian direction of our supercell. In the case of the $\bar{6}$ point group, there are two independent linear electro-optic constants as the tensor has the following expression:

$$r_{ijk} = \begin{pmatrix} r_{11} & -r_{11} & 0 & 0 & 0 & -r_{22} \\ -r_{22} & r_{22} & 0 & 0 & 0 & -r_{11} \\ 0 & 0 & 0 & 0 & 0 & 0 \end{pmatrix}, \tag{53}$$

Using the previous method, we find the following rescaling equations for the electro-optic constants of the 2D layer:

$$r_{11}^{2D} = \frac{c}{t}\left(\frac{\varepsilon_{11}^{SC}}{\varepsilon_{11}^{2D}}\right)^2 r_{11}^{SC} \tag{54}$$

$$r_{22}^{2D} = \frac{c}{t}\left(\frac{\varepsilon_{22}^{SC}}{\varepsilon_{22}^{2D}}\right)^2 r_{22}^{SC} \tag{55}$$

*Point group 622*

The 622 point group has only one independent electro-optic constant, as

$$r_{ijk} = \begin{pmatrix} 0 & 0 & 0 & r_{14} & 0 & 0 \\ 0 & 0 & 0 & 0 & -r_{14} & 0 \\ 0 & 0 & 0 & 0 & 0 & 0 \end{pmatrix}. \tag{56}$$

The rescaling of the electro-optic constants to obtain the "real" EO constant of the 2D layer is

$$r_{14}^{2D} = \frac{\varepsilon_{11}^{SC}\varepsilon_{33}^{SC}}{\varepsilon_{11}^{2D}\varepsilon_{33}^{2D}} r_{14}^{SC} \tag{57}$$

*Point group 6mm*

Directing the 6-fold symmetry axis along the third Cartesian direction of our supercell too, the EO tensor has the form:

$$r_{ijk} = \begin{pmatrix} 0 & 0 & 0 & 0 & r_{24} & 0 \\ 0 & 0 & 0 & r_{24} & 0 & 0 \\ r_{31} & r_{31} & r_{33} & 0 & 0 & 0 \end{pmatrix}. \tag{58}$$

Using the methods described above in the case of the *mm2* point group, the rescaling laws are the following:

$$r_{33}^{2D} = \frac{c}{t}\frac{\varepsilon_{33}^{2D}}{\varepsilon_{33}^{SC}} r_{33}^{SC} \tag{59}$$

$$r_{31}^{2D} = \frac{c}{t}\frac{\varepsilon_{33}^{2D}}{\varepsilon_{33}^{SC}}\left(\frac{\varepsilon_{11}^{SC}}{\varepsilon_{11}^{2D}}\right)^2 r_{31}^{SC} \tag{60}$$

$$r_{24}^{2D} = \frac{\varepsilon_{22}^{SC}\varepsilon_{33}^{SC}}{\varepsilon_{22}^{2D}\varepsilon_{33}^{2D}} r_{24}^{SC} \tag{61}$$

*Point group 6*

The 6-fold axis is assumed to lie in the third Cartesian direction of our supercell, as well. The EO tensor has the following symmetry:

$$r_{ijk} = \begin{pmatrix} 0 & 0 & 0 & r_{14} & r_{24} & 0 \\ 0 & 0 & 0 & r_{24} & -r_{14} & 0 \\ r_{31} & r_{31} & r_{33} & 0 & 0 & 0 \end{pmatrix}. \tag{62}$$

The renormalization to be carried out to obtain the 2D EO coefficient from those calculated by DFPT in a supercell is:

$$r_{31}^{2D} = \frac{c}{t} \frac{\varepsilon_{33}^{2D}}{\varepsilon_{33}^{SC}} \left(\frac{\varepsilon_{11}^{SC}}{\varepsilon_{11}^{2D}}\right)^2 r_{31}^{SC} \tag{63}$$

$$r_{33}^{2D} = \frac{c}{t} \frac{\varepsilon_{33}^{2D}}{\varepsilon_{33}^{SC}} r_{33}^{SC} \tag{64}$$

$$r_{14}^{2D} = \frac{\varepsilon_{11}^{SC} \varepsilon_{33}^{SC}}{\varepsilon_{11}^{2D} \varepsilon_{33}^{2D}} r_{14}^{SC} \tag{65}$$

$$r_{24}^{2D} = \frac{\varepsilon_{22}^{SC} \varepsilon_{33}^{SC}}{\varepsilon_{22}^{2D} \varepsilon_{33}^{2D}} r_{24}^{SC} \tag{66}$$

### E. Trigonal point groups

For all trigonal point groups, we assume that the 3-fold rotation axis is directed along the third Cartesian direction.

*Point group* $3m$

The EO tensor symmetry is

$$r_{ijk} = \begin{pmatrix} 0 & 0 & 0 & 0 & r_{24} & -r_{22} \\ -r_{22} & r_{22} & 0 & r_{24} & 0 & 0 \\ r_{31} & r_{31} & r_{33} & 0 & 0 & 0 \end{pmatrix}. \tag{67}$$

The renormalization to apply to deduce the 2D EO coefficients from those calculated by DFPT in a supercell are

$$r_{31}^{2D} = \frac{c}{t} \frac{\varepsilon_{33}^{2D}}{\varepsilon_{33}^{SC}} \left(\frac{\varepsilon_{11}^{SC}}{\varepsilon_{11}^{2D}}\right)^2 r_{31}^{SC} \tag{68}$$

$$r_{33}^{2D} = \frac{c}{t} \frac{\varepsilon_{33}^{2D}}{\varepsilon_{33}^{SC}} r_{33}^{SC} \tag{69}$$

$$r_{22}^{2D} = \frac{c}{t} \left(\frac{\varepsilon_{22}^{SC}}{\varepsilon_{22}^{2D}}\right)^2 r_{22}^{SC} \tag{70}$$

$$r_{24}^{2D} = \frac{\varepsilon_{22}^{SC} \varepsilon_{33}^{SC}}{\varepsilon_{22}^{2D} \varepsilon_{33}^{2D}} r_{24}^{SC} \tag{71}$$

*Point group* $32$

The EO tensor symmetry is

$$r_{ijk} = \begin{pmatrix} r_{11} & -r_{11} & 0 & r_{14} & 0 & 0 \\ 0 & 0 & 0 & 0 & -r_{14} & -r_{11} \\ 0 & 0 & 0 & 0 & 0 & 0 \end{pmatrix}. \tag{72}$$

The renormalization is performed through the following relations:

$$r_{11}^{2D} = \frac{c}{t} \left(\frac{\varepsilon_{11}^{SC}}{\varepsilon_{11}^{2D}}\right)^2 r_{11}^{SC} \tag{73}$$

$$r_{14}^{2D} = \frac{\varepsilon_{11}^{SC} \varepsilon_{33}^{SC}}{\varepsilon_{11}^{2D} \varepsilon_{33}^{2D}} r_{14}^{SC} \tag{74}$$

*Point group* $3$

The EO tensor symmetry is

$$r_{ijk} = \begin{pmatrix} r_{11} & -r_{11} & 0 & r_{14} & r_{24} & -r_{22} \\ -r_{22} & r_{22} & 0 & r_{24} & -r_{14} & -r_{11} \\ r_{31} & r_{31} & r_{33} & 0 & 0 & 0 \end{pmatrix}. \tag{75}$$

The renormalization relationships can be written as:

$$r_{11}^{2D} = \frac{c}{t}\left(\frac{\varepsilon_{11}^{SC}}{\varepsilon_{11}^{2D}}\right)^2 r_{11}^{SC} \tag{76}$$

$$r_{31}^{2D} = \frac{c}{t}\frac{\varepsilon_{33}^{2D}}{\varepsilon_{33}^{SC}}\left(\frac{\varepsilon_{11}^{SC}}{\varepsilon_{11}^{2D}}\right)^2 r_{31}^{SC} \tag{77}$$

$$r_{33}^{2D} = \frac{c}{t}\frac{\varepsilon_{33}^{2D}}{\varepsilon_{33}^{SC}} r_{33}^{SC} \tag{78}$$

$$r_{22}^{2D} = \frac{c}{t}\left(\frac{\varepsilon_{22}^{SC}}{\varepsilon_{22}^{2D}}\right)^2 r_{22}^{SC} \tag{79}$$

$$r_{14}^{2D} = \frac{\varepsilon_{11}^{SC}\varepsilon_{33}^{SC}}{\varepsilon_{11}^{2D}\varepsilon_{33}^{2D}} r_{14}^{SC} \tag{80}$$

$$r_{24}^{2D} = \frac{\varepsilon_{22}^{SC}\varepsilon_{33}^{SC}}{\varepsilon_{22}^{2D}\varepsilon_{33}^{2D}} r_{24}^{SC} \tag{81}$$

**F. Cubic point groups**

Note that cubic point groups for the supercell can only arise in very specific cases. As one may want to vary the vacuum thickness to check proper convergence of the EO coefficients, it is unlikely that the supercell will be cubic in most cases.

*Point group $\bar{4}3m$ and $23$*
The EO tensor symmetry is

$$r_{ijk} = \begin{pmatrix} 0 & 0 & 0 & r_{14} & 0 & 0 \\ 0 & 0 & 0 & 0 & r_{14} & 0 \\ 0 & 0 & 0 & 0 & 0 & r_{14} \end{pmatrix}. \tag{82}$$

The rescaling relation to obtain the 2D EO coefficients is

$$r_{14}^{2D} = \frac{\varepsilon_{11}^{SC}\varepsilon_{33}^{SC}}{\varepsilon_{11}^{2D}\varepsilon_{33}^{2D}} r_{14}^{SC} \tag{83}$$

**G. DFT calculations**

Let us now conduct first-principles calculations to confirm the validity of these formula. Here, we chose to investigate the 2D ferroelectric SnSe monolayer. Figure 2 shows its crystal structure, which has the $Pmn2_1$ space group, and thus the $mm2$ point group. The spontaneous polarization in monolayer SnSe is along the in-plane x direction. Different vacuum regions, with the corresponding thickness of the whole supercell being denoted by $c$ as consistent with Fig. 1 (with $c$ being here at least 15 Å) along the out-of-plane z direction, are introduced in a slab structure to avoid the interaction between the neighboring periodic images. $t$ represents the effective thickness of the SnSe monolayer, and is practically taken to be equal to 5.7 Å here.

To verify the validity of Eqs. (16), (17), (19), (22) and (23) within the $mm2$ point group, we first calculate the so-called *clamped* EO coefficients in monolayer SnSe in its $Pmn2_1$ ground state for the aforementioned slab geometries using the technique of Ref[9]. Note that clamped EO coefficients correspond to strain-free mechanical boundary conditions, and include electronic and ionic contributions but neglect any modification of the unit cell shape. Figure 3(a) shows the results, ($r_{11}^{\eta,SC}, r_{12}^{\eta,SC}, r_{13}^{\eta,SC}, r_{35}^{\eta,SC}$ and $r_{26}^{\eta,SC}$), directly arising from this DFPT technique for different $c$ thickness along the out-of-plane z direction. Figure 3(b) then displays the corresponding clamped EO coefficients of the 2D material obtained from the use of Eqs. (16), (17), (19), (22) and (23), as well as, Eqs. (6) and (7) to extract the 2D dielectric tensor elements from the "SC" ones directly obtained by DFT for the slabs. One can clearly see that the EO coefficients of the supercell strongly depend on the c supercell lattice constant along the out-of-plane direction. In contrast, the renormalized EO coefficients in the 2D material of Fig. 3(b) no longer depend on $c$ when using the rescaling equations for the EO constants of the 2D layer, which therefore attests to the validity of the formula derived above for EO coefficient in 2D materials. One can also realize that the magnitude of these coefficients is really different between the ones associated with the slab and those corresponding to the 2D material. For instance, the clamped EO coefficients of the supercell $r_{13}^{\eta,SC}$ and $r_{26}^{\eta,SC}$ are $-1.5$ and 865.4 pm/V, respectively, when $c$ is equal to about 70 Å. On the other hand, the renormalized EO coefficients of the 2D material $r_{13}^{\eta,2D}$ and $r_{26}^{\eta,2D}$ are, in magnitude, larger by a factor of 12 and smaller by a factor 6, respectively, since they are equal to about $-18.0$ and 142.2 pm/V, respectively, for different $c$ thickness. Note that the EO coefficients originate from different electronic and ionic contributions (corresponding to various phonon modes), which results in different sign and magnitude of EO coefficients.

Moreover, Figs. 3(c) and 3(d) show the clamped EO coefficients $r_{11}^{\eta,SC}$ and $r_{12}^{\eta,SC}$ for the supercell and 2D material, as a function of electric field for two different $c$ thickness, respectively. We found that the EO coefficients of the 2D material versus electric field are also independent on the $c$ thickness of the supercell, which once again attests to the validity of the formula derived for EO coefficient in 2D materials.

Let us now present some EO results for monolayer BN. The monolayer BN has a $P\bar{6}m2$ space group ($\bar{6}m2$ point group) and the EO tensor has one independent element, $r_{22}$ [see Eq. (51)]. Figure 4(a) shows that the in-plane and out-of-plane elements of the electronic dielectric tensor strongly depends on the $c$ thickness of the supercell in monolayer BN. In contrast, Fig. 4(b) demonstrates that the renormalized dielectric constants are no longer dependent on the $c$ thickness when using Eqs. (6) and (7) and choosing an effective monolayer thickness $t = 3.4$ Å [see the inset of Fig. 4(b)]. Note that the thickness in monolayer BN is very close to the previous theoretical value of 3.17 Å[31]. We also checked the EO coefficients for the supercell and 2D material as a function of the $c$ thickness of the supercell in monolayer BN. Figure 4(c) shows that the clamped EO coefficient $r_{22}^{\eta,SC}$ obviously depends on the $c$ thickness. In sharp contrast, the EO coefficient of the 2D material of Fig. 4(d) no longer relies on the $c$ thickness when using the formula of Eq. (52), implying once again the validity of the derived EO coefficient formula in 2D materials.

Note that our method is also valid for unclamped EO coefficients. More precisely, the unclamped EO coefficients need to add contributions associated with elasto-optic and piezoelectric coefficients from the clamped ones. The piezoelectric strain coefficients are independent on the $c$ thickness, while the elasto-optic coefficients need to be rescaled in the same way as the EO coefficients.

## IV. SUMMARY

In summary, DFPT calculations employing periodic boundary conditions introduce spurious effects when dealing with 2D materials. Beyond the obvious periodic image interaction which should be limited as much as possible, the presence of vacuum in a supercell geometry artificially modifies the dielectric and electro-optic response calculated. The present paper shows a method to renormalize the electro-optic coefficients obtained by DFPT with periodic boundary conditions. The renormalization relationships to recover the "true" 2D electro-optic coefficients are derived for the appropriate cubic, tetragonal, orthorhombic, hexagonal and trigonal point groups. These renormalizations are further numerically confirmed by conducting DFT calculations. It is also important to realize that these renormalizations are general to any method that attempts at deriving the EO coefficients from a slab calculation. In other words, they are not only restricted to DFPT.

## ACKNOWLEDGEMENTS

This work is supported by the National Natural Science Foundation of China (Grants No. 12374092 and No. 11804138), Natural Science Basic Research Program of Shaanxi (Program No. 2023-JC-YB-017), Shaanxi Fundamental Science Research Project for Mathematics and Physics (Grant No. 22JSQ013), "Young Talent Support Plan" of Xi'an Jiaotong University (Grant No. WL6J004), the Open Project of State Key Laboratory of Surface Physics (Grant No. KF2023_06), the Fundamental Research Funds for the Central Universities, and the HPC Platform of Xi'an Jiaotong University. H.X. acknowledges the support from the Ministry of Science and Technology of the People's Republic of China (No. 2022YFA1402901) and NSFC (Grants No. 11825403, 11991061, and 12188101). L.B. thanks the MonArk NSF Quantum Foundry supported by the National Science Foundation Q-AMASE-i Program under NSF Award No. DMR-1906383, ARO Grant No. W911NF-21-1-0113 and the Vannevar Bush Faculty Fellowship (VBFF) Grant No. N00014-20-1-2834 from the Department of Defense.

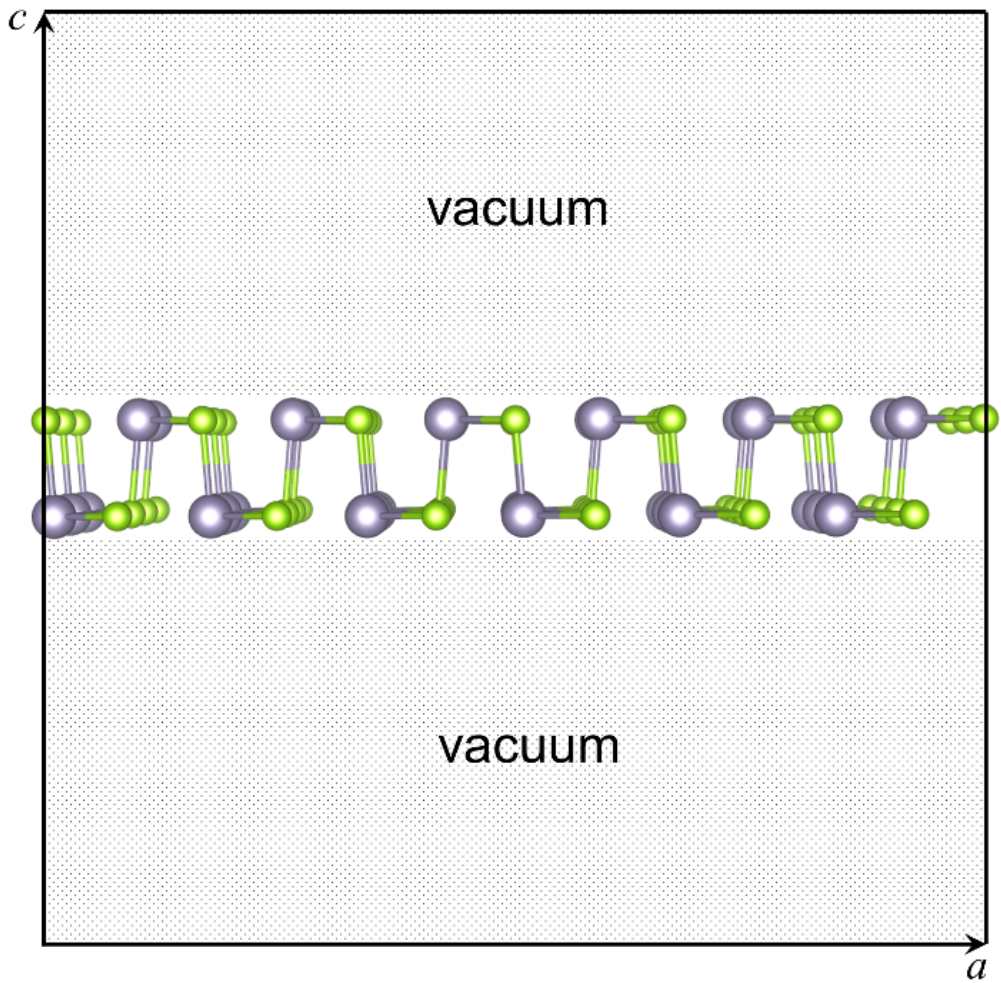

**Fig. 1** Typical slab calculation of a 2D material in DFT codes using periodic boundary conditions. The supercell $c$ lattice constant is much larger than the effective thickness of the 2D material.

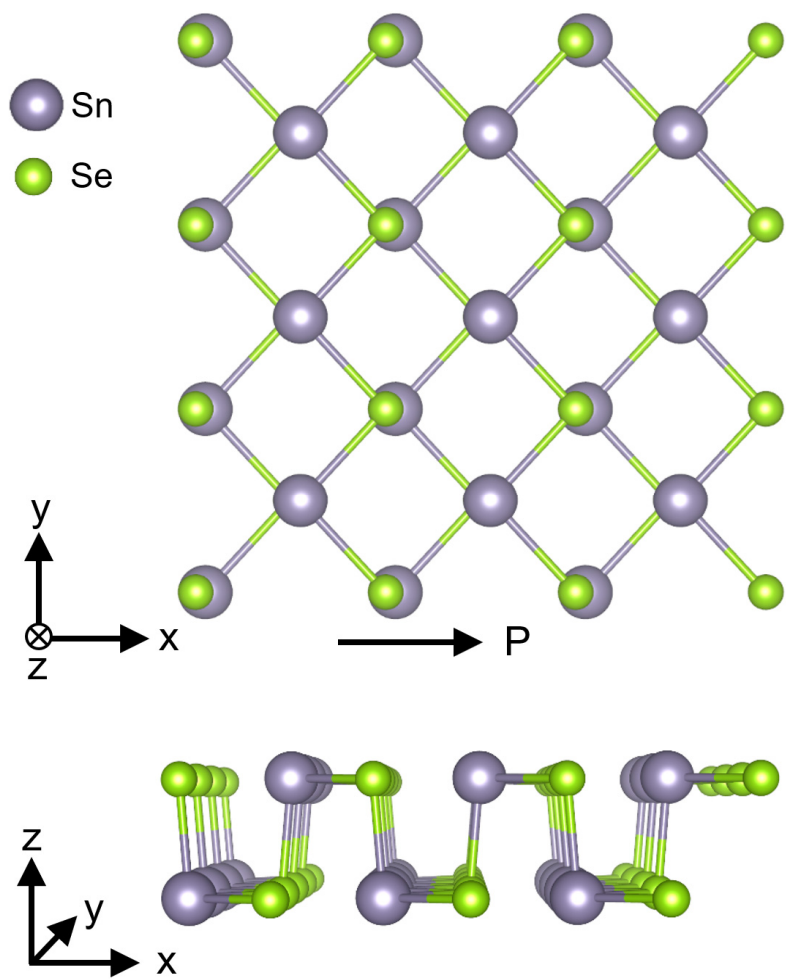

**Fig. 2** The top and side views of crystal structures in monolayer SnSe.

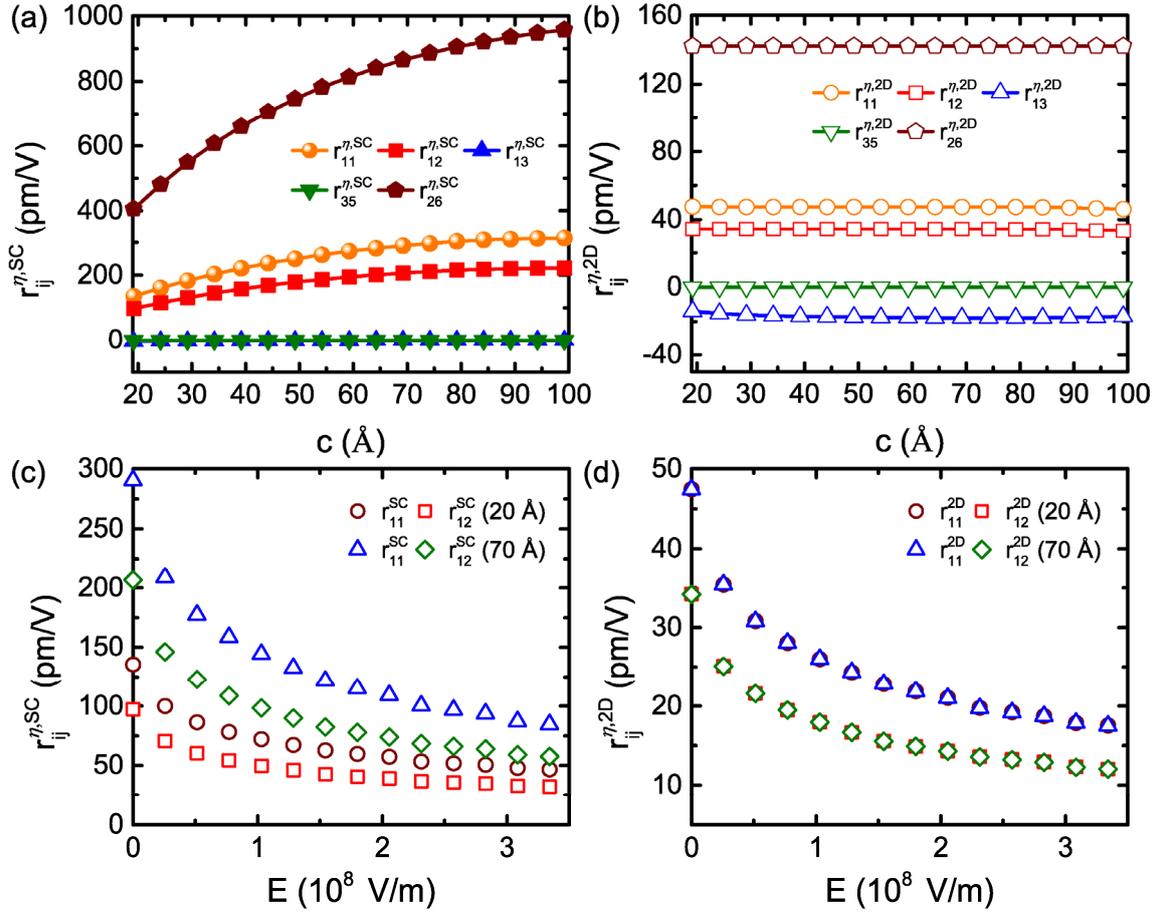

**Fig. 3** Clamped EO coefficients for (a) the supercell and (b) 2D material versus different supercell lattice constant $c$ along the out-of-plane z direction in monolayer SnSe. EO coefficients for (c) the supercell and (d) 2D material as a function of the electric field for two different $c$ thickness of the supercell in monolayer SnSe.

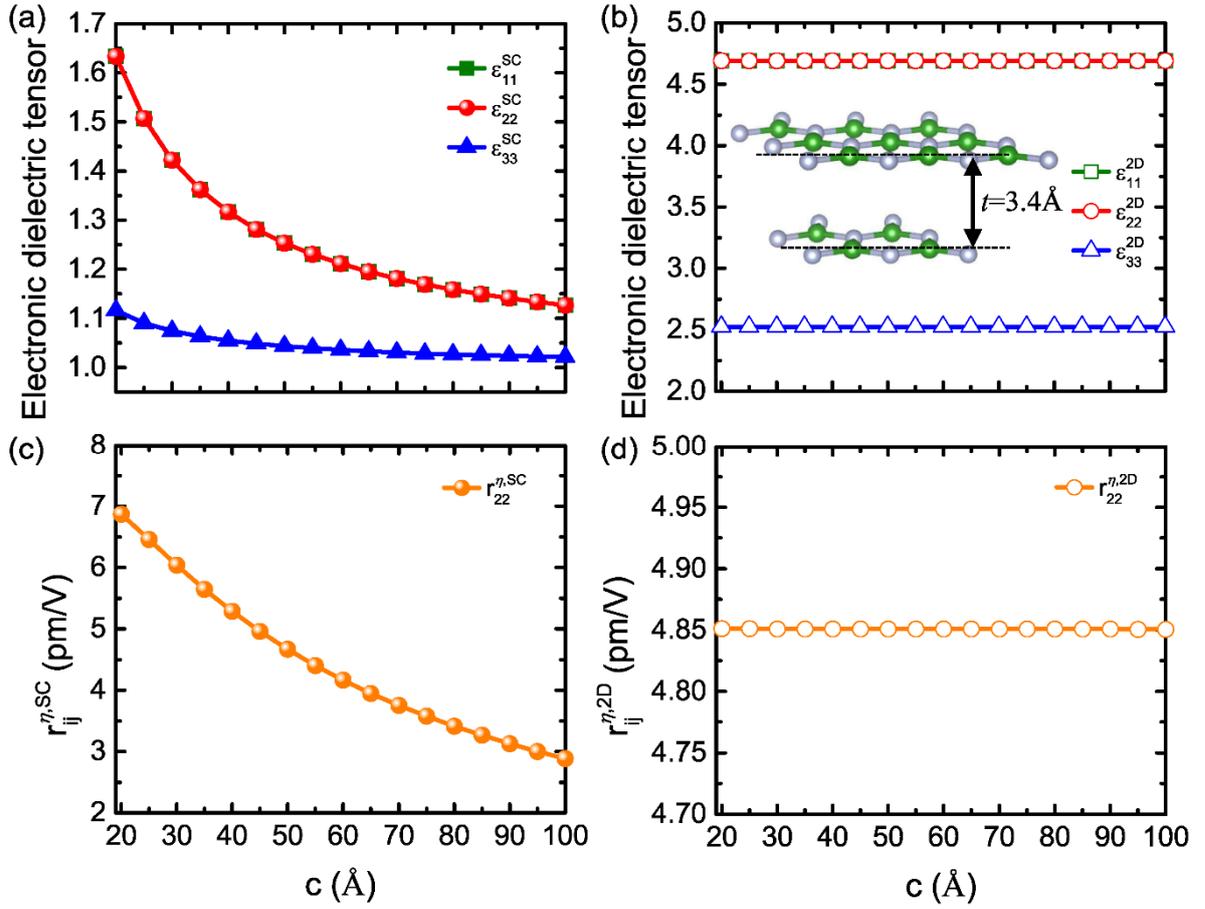

**Fig. 4** Electronic dielectric tensor for (a) the supercell and (b) 2D material as a function of the $c$ thickness of the supercell in monolayer BN. EO coefficients for (c) the supercell and (d) 2D material versus the $c$ thickness of the supercell in monolayer BN. The inset of (b) shows the distance between the center of the top and bottom planes in the bilayer BN, which corresponds to the thickness $t$ of the monolayer BN.